\documentclass[12pt,a4paper]{article}
\usepackage{epsfig}
\usepackage{amsmath}
\usepackage{amssymb}
\usepackage{subfigure}
\newcommand{\spp}{\vphantom{\bigg(}}

\def\gev{{\rm GeV}}

\begin{document}
\title{Rare decays $B\rightarrow M\nu\bar{\nu}$
in the $TC2$ model \\ and the $LHT$ model
 \hspace*{-0.8cm}  }

\author{Chong-Xing Yue, Jiao Zhang, Wei Liu\\
{\small  Department of Physics, Liaoning Normal University, Dalian
116029, China}\thanks{cxyue@lnnu.edu.cn}\\}
\date{}

\maketitle
%---------------------------------------------------------------------------
\begin{abstract}
In the framework of the topcolor-assisted technicolor ($TC2$) model
and the littlest Higgs model with $T$-parity ($LHT$ model), we
consider the rare $B$ decays $B\rightarrow M\nu\bar{\nu}$ with
$M=\pi,K,\rho,$ or $K^{\ast}$. We find that the contributions of the
$TC2$ model to the branching ratios of these decay processes are
larger than those for the $LHT$ model. The experimental upper limits
for some branching ratios can give severe constraints on the free
parameters of the $TC2$ model.

\vspace{1cm}

PACS number: 12.60.Cn, 12.60.Fr, 13.20.He
\end{abstract}

\newpage
%%%%%%%%%%%%%%%%%%%%%%

\noindent{\bf \large 1. Introduction}

$B$ physics plays an important role in testing the standard model
($SM$) and $B$ decays are sensitive to new physics beyond the $SM$
[1]. A large number of $B$ mesons is produced in $B$ factories, such
as $Belle$ and $BaBar$ experiments. Recently, there has been
significant experimental improvement in  measurements of the flavor
changing neutral current ($FCNC$) processes related to $B$ mesons at
$B$ factories and Tevatron. $B$ physics is entering the era of
precision measurement, which is not far from possibly revealing new
physics. Rare $B$ decays, which are mediated by $FCNCs$, are good
places for new physics to enter through exchange of new particles at
the loop level or through new interactions at the tree level. Thus,
studying rare $B$ decays in some specific new physics models is very
interesting and needed.

The rare $B$ decays $B\rightarrow M \nu\bar{\nu}$ with $M= \pi, K,
\rho$, or $ K^{\ast}$  belong to the theoretically cleanest decays
in the field of $FCNC$ processes, due to the absence of photonic
penguin contributions and strong suppression of light quark
contributions. Since they are significantly suppressed by the loop
momentum and off-diagonal Cabibbo-Kobayashi-Maskawa ($CKM$)
matrix-elements in the $SM$ and their long-distance contributions
are generally subleading, these rare decay processes are considered
as excellent probes of new physics beyond the $SM$. This fact has
lead to lot of works for studying the contributions of some popular
new physics models to the rare decays $B\rightarrow M \nu\bar{\nu}$
in Refs. [2, 3, 4].

In spite of the above theoretical advantages, experimental search of
the rare decays $B\rightarrow M \nu\bar{\nu}$ is a hard task.
However, with the advent of Super-B facilities [5], the prospects of
measuring the branching ratios of the rare decay processes
$B\rightarrow M \nu\bar{\nu}$ in the next decade are possible and it
seems appropriate to further study these decays in order to motivate
more experimental efforts to measure their branching ratios and
related observables. So, in this paper, we reconsider these rare
decays processes in the context of the topcolor-assisted technicolor
($TC2$) model [6] and the littlest Higgs model [7, 8] with $T$ parity
(called $LHT$ model) [9]. Our numerical results show that these two
kinds of popular new physics models can indeed give significant
contributions to these rare $B$ decay processes and the current
experimental limits for some of these processes can put severe
constraints on the free parameters of the $TC2$ model. Furthermore,
in the context of the $TC2$ model, we consider that the
contributions of the nonuniversal gauge boson $Z'$ to the quark
level transition processes $b \to sl^+l^-$ are correlated with those
for the quark level transition processes $b \to s\nu\bar{\nu}$ and
recalculate the contributions of $Z'$ to the rare decay processes
$B\rightarrow M \nu\bar{\nu}$. We also compare our numerical results
with those obtained in Ref. [4].

In the rest of this paper, we will give our results in detail. After
briefly summarizing the essential features of the $TC2$ model, we
calculate the contributions of the new particles predicted by this
model to the rare $B$ decays $B\rightarrow M \nu\bar{\nu}$ with $M=
\pi, K, \rho, K^{\ast}$ in section 2. To compare our results
obtained in the context of the $TC2$ model with those of the $LHT$
model, the branching ratios of these decay processes contributed by
the $LHT$ model are estimated in section 3 by using the results of
Refs. [10, 11]. Our conclusions are given in section 4.

\vspace{0.3cm}

\noindent{\bf \large 2. The $TC2$ model and the rare decays $B\rightarrow M \nu\bar{\nu}$}

In the $TC2$ model [6], topcolor interactions, which are not
flavor-universal and mainly couple to the third generation fermions,
generally generate small contributions to electroweak symmetry
breaking ($EWSB$) and give rise to the main part of the top quark
mass. Thus, the nonuniversal gauge boson $Z'$ has large Yukawa
couplings to the third generation fermions. Such features can result
in large tree level flavor changing ($FC$) couplings of the
nonuniversal gauge boson $Z'$ to ordinary fermions when one writes
the interaction in the fermion mass eigen-basis.

The flavor diagonal ($FD$) couplings of the nonuniversal gauge boson
$Z'$ to ordinary fermions, which are related  to our calculation,
can be written as [6, 12, 13]:
\begin{eqnarray}
\mathcal{L}^{FD}_{Z'}& = & -\sqrt{4\pi K_{1}}\left\{Z'_{\mu}\left[
\frac{1}{6}\bar{t}_{L}\gamma^{\mu}t_{L}+\frac{1}{6}\bar{b}_{L}
\gamma^{\mu}b_{L}+\frac{2}{3}\bar{t}_{R}\gamma^{\mu}t_{R}-\frac{1}{3}\bar{b}_{R}\gamma^{\mu}b_{R}-\frac{1}{2}\bar{\nu}_{\tau L}
\gamma^{\mu}\nu_{\tau L}\right]\right.\nonumber\\
&&-\left.\tan^{2}\theta'Z'_{\mu}\left[
\frac{1}{6}\bar{s}_{L}\gamma^{\mu}s_{L}-\frac{1}{3}\bar{s}_{R}\gamma^{\mu}s_{R}+\frac{1}{6}\bar{d}_{L}\gamma^{\mu}d_{L}-
\frac{1}{3}\bar{d}_{R}\gamma^{\mu}d_{R}
-\frac{1}{2}\bar{\nu}_{\mu L}\gamma^{\mu}\nu_{\mu L}\right.\right.\nonumber\\
&&-\left.\left.\frac{1}{2}\bar{\nu}_{e L}\gamma^{\mu}
\nu_{e L}\right]\right\},
\end{eqnarray}
where $K_{1}$ is the coupling constant, $\theta'$ is the mixing
angle with $\tan\theta'=\frac{g_{1}}{\sqrt{4\pi K_{1}}}$ , and
$g_{1}$ is the ordinary hypercharge gauge coupling constant.

The $FC$ couplings of the nonuniversal gauge boson $Z'$ to down-type
quarks, i.e. $Z'bd_{j}$ with $j=s$ or
 $d$,  can be written as [13]:
\begin{equation}\label{eq1}
\mathcal {L}^{FC}_{Z'}=
-\frac{g_{1}}{2}\cot\theta'Z'^{\mu}\left\{\frac{1}{3}D^{bb}_{L}D^{bd_{j}\ast}_{L}\bar{d_{j}}_{L}\gamma_{\mu}b_{L}
-\frac{2}{3}D^{bb}_{R}D^{bd_{j}\ast}_{R}\bar{d_{j}}_{R}\gamma_{\mu}b_{R}+
h.c.\right\},
\end{equation}
where $D_L$ and $D_R$ are matrices which rotate the down-type left-
and right- handed quarks from the quark field to mass eigen-basis,
respectively.

\vspace{0.8cm}
\begin{figure}[htb]
\begin{center}
\epsfig{file=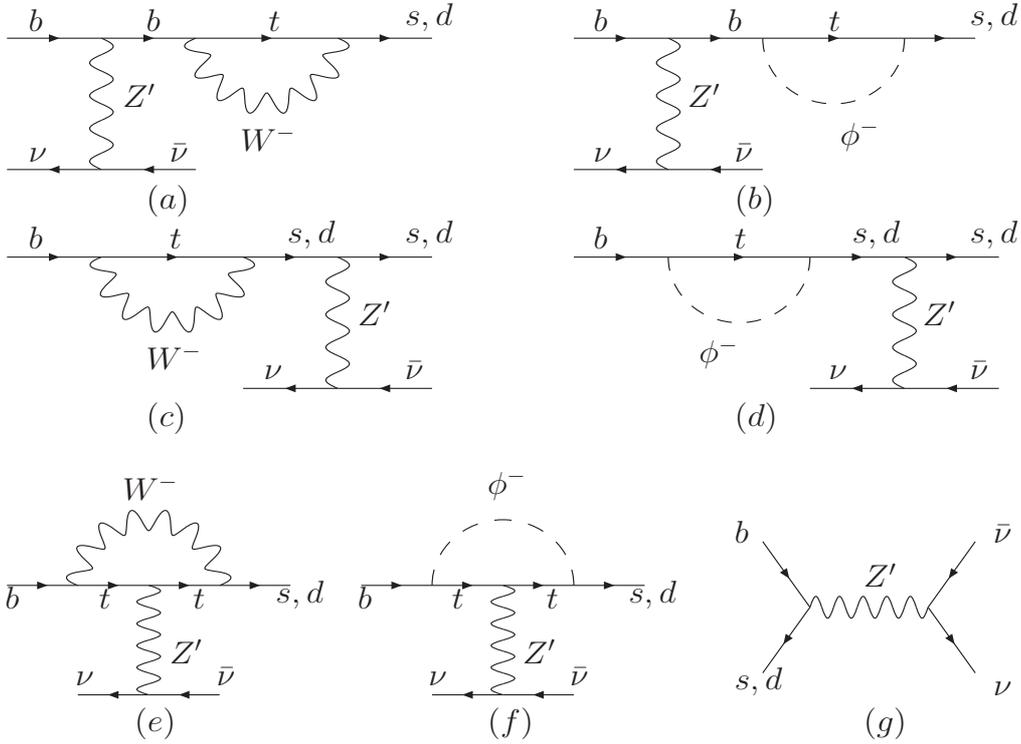,width=380pt,height=280pt}
 \vspace{-0.3cm}\caption{The penguin and tree level
diagrams for $Z'$ contributing to the rare decay \hspace*{1.8cm}
processes $b\rightarrow (s,d)\nu\bar{\nu}$ in
the $TC2$ model.}
 \label{ee}
\end{center}
\vspace{-0.5cm}
\end{figure}

The quark level transition $b\rightarrow d_{j}\nu\bar{\nu}$ ($j=s$
or $d$) is responsible for the semi-leptonic decays $B\rightarrow
M\nu\bar{\nu}$($M=\pi,K,\rho,K^{\ast}$). From the above discussions,
we can see that the nonuniversal gauge boson $Z'$ can contribute to
the rare decay processes $b\rightarrow (s,d)\nu\bar{\nu}$ at the
tree level and the one loop level. The relevant Feynman diagrams are
shown in Fig.1. In these diagrams, the Goldstone boson $\phi$ is
introduced by the 't Hooft-Feynman gauge, which can cancel the
divergence in self-energy diagrams.

The effective Hamiltonian for the transition $b \rightarrow
d_{j}\nu_{i}\bar{\nu_{i}}$ ($j= s,d$ and $i = e,\mu,\tau $)
 can be written as [14]:
\begin{eqnarray}
\mathcal{H}_{eff}(b\rightarrow d_{j}\nu_{i}\bar{\nu_{i}})& = & C^{\nu}_{L}\bar{b}\gamma_{\mu}(1-\gamma_{5})d_{j}\bar{\nu_{i}}
\gamma^{\mu}(1-\gamma_{5})\nu_{i}
+C^{\nu}_{R}\bar{b}\gamma_{\mu}(1+\gamma_{5})d_{j}\bar{\nu_{i}}\gamma^{\mu}
(1-\gamma_{5})\nu_{i}\nonumber\\
&&\equiv C^{\nu}_{L}\mathcal{O}_{L}+C^{\nu}_{R}\mathcal{O}_{R}
\end{eqnarray}
$\mathcal{O}_{L(R)}$ and $C^{\nu}_{L(R)}$ represent the left-
(right-) handed operators and the corresponding coefficients,
respectively. By the way, these operators and coefficients are
defined with opposite signs w.r.t. those in Ref. [4]. In the $SM$,
the processes $b\rightarrow d_{j}\nu_{i}\bar{\nu_{i}}$ proceed via
$W$ box and $Z$ penguin diagrams, therefore only purely left-handed
currents $\bar{b}\gamma_{\mu}(1-\gamma_{5})d_{j}\bar{\nu_{i}}
\gamma^{\mu}(1-\gamma_{5})\nu_{i}$ are present. The corresponding
left-handed coefficient reads
\begin{eqnarray}
C^{\nu}_{L,SM}=\frac{G_{F}\alpha_{e}}{2\pi\sqrt{2}}
V_{td_{j}}V_{tb}^{\ast}\frac{X(x_{t})}{\sin^{2}\theta_{w}},
\end{eqnarray}
where $G_{F}$ is the Fermi constant, $\alpha_{e}$ is the fine
structure constant, $\theta_{w}$ is the Weinberg angle, and $V_{ij}$
is the $CKM$ matrix element. The $SM$ Inami-Lim function $X(x_{t})$
[15] is dominated by the short distance dynamics associated with top
quark exchange. $\mathcal{O}_{R}$ is one new right-handed operator
induced by new physics effects with $C^{\nu}_{R}$ only receiving
contributions from new physics beyond the $SM$.

It is obvious that the nonuniversal gauge boson $Z'$ predicted by
the $TC2$ model can give corrections to the coefficient
$C^{\nu}_{L,SM}$ via both the penguin and  tree level diagrams,
while it can give contributions to the coefficient $C^{\nu}_{R}$
only via the tree level diagram. From the Feynman diagrams given in
Fig.1, we can obtain the corresponding coefficients in Eq.(3)
contributed by $Z'$. For $\nu_{i}= \nu_{e}$ and  $\nu_{\mu}$, their
expression forms can be written as:
\begin{eqnarray}
C^{\nu}_{L}  &=& C^{\nu}_{L,SM}+\frac{\tan^{2}\theta'g^{2}}{4M_{Z'}}
\left[\frac{g_{2}^{2}V^{\ast}_{tb}V_{td_{j}}}{8(4\pi^{2})}X^{TC}(x_{t})
-\frac{1}{12}D^{bb^{\ast}}_{L}D^{bd_{j}}_{L}\right],\\
C^{\nu}_{R}   &=&  \frac{\tan\theta'g^{2}}{24M_{Z'}}D^{bb^{\ast}}_{R}D^{bd_{j}}_{R},\\
X^{TC}(x_{t}) &=& C_{a}(x_{t})+ C_{b}(x_{t})+ C_{c}(x_{t}).
\end{eqnarray}
Here $x_{t}=\frac{m_{t}^{2}}{M_{W}^{2}}$ and $g=\sqrt{4\pi K_{1}}$.
Using the method given in Ref. [15], we can calculate the functions
$C_{a}(x_{t})$, $C_{b}(x_{t})$ and $C_{c}(x_{t})$ in the framework
of the $TC2$ model. $C_{a}(x_{t})$ is obtained from the penguin
diagrams Fig.1 $(a)$,$(b)$,$(c)$ and $(d)$, $C_{b}(x_{t})$ is
obtained from the penguin diagram Fig.1 $(e)$, $C_{c}(x_{t})$ is
obtained from the penguin diagram Fig.1 $(f)$. The third term of the
coefficient $C^{\nu}_{L}$ and the coefficient $C^{\nu}_{R}$ are
contributed by Fig.1 $(g)$. The detailed expression forms of these
functions are listed in Appendix.

From Eq. (2) we can see that, for the processes $b\rightarrow
d_{j}\nu_{\tau}\bar{\nu}_{\tau}$, the expression forms of the
coefficients $C^{\nu}_{L}$ and  $C^{\nu}_{R}$ are similar to those
for the processes $b\rightarrow d_{j}\nu_{e}\bar{\nu}_{e}$. However,
the factor $\tan^{2}\theta'$ should be omitted.

The decay amplitudes of  the exclusive semi-leptonic decay processes
$B\rightarrow M\nu\bar{\nu}$($M=\pi,K,\rho,K^{\ast}$) can be
obtained after evaluating matrix elements of the quark operators
given in Eq. (3) between the initial $|B>$ and final $|M>$ states.
The hadronic matrix elements for $B\rightarrow P$ decay ($P$ is a
pseudoscalar meson, $\pi$ or $K$) can be parameterized in terms of
the form factors $f^{P}_{+}(s_{B})$ and $f^{P}_{0}(s_{B})$ as [2, 3,
4, 16]:
\begin{equation}
c_{p}\langle
P(p)|\bar{u}\gamma_{\mu}b|B(p_{B})\rangle=f^{P}_{+}(s_{B})(p+p_{B})_{\mu}+
[f^{P}_{0}(s_{B})-f^{P}_{+}(s_{B})]\frac{m^{2}_{B}-m^{2}_{P}}{s_{B}}q_{\mu},
\end{equation}
where the factor $c_{P}$ accounts for flavor content of particles
($c_{P}$ = $\sqrt{2}$ for $\pi^{0}$ and $c_{P}$ = 1 for $\pi^{-}$,
$K^{-}$) and $s_{B}$ = $q^{2}$ ($q$ = $p_{B} - p$ = $p_{\nu} +
p_{\bar{\nu}}$).  For $B\rightarrow V$ decay ($V$ is a vector mesons
$K^{\ast}$ or $\rho$ ), its hadronic matrix elements can be written
in terms of five form factors:
\begin{eqnarray}
c_V\langle
V(p,\varepsilon^{\ast})|\bar{u}\gamma_{\mu}(1-\gamma_5)b|B
(p_{_{B}})\rangle &&=\frac{2V(s_B)}{m_B+m_V}
\epsilon_{\mu\nu\alpha\beta}\varepsilon^{\ast\nu}p_{_B}^{\alpha}p^{\beta}\nonumber\\
&&-i\left[\varepsilon_{\mu}^\ast(m_B+m_V)A_1(s_B)
-(p_{_B}+p)_{\mu}({\varepsilon^\ast}\cdot{p_{_B}})\frac{A_2(s_B)}
{m_B+m_V}\right]\nonumber \\
&&+iq_{\mu}({\varepsilon^\ast}\cdot{p_{_B}})\frac{2m_V}{s_B}
[A_3(s_B)-A_0(s_B)]
\end{eqnarray}
with
\begin{eqnarray}
V(s_{B})&=& \frac{V(0)}{1-s_{B}^{2}/M^{2}_{p}},\quad\quad M_{p} = 5~\rm  GeV,\\
A_3(s_B)&=&\frac{m_B+m_V}{2m_V}A_1(s_B)-\frac{m_B-m_V}{2m_V}A_2(s_B),
\end{eqnarray}
where $c_V=\sqrt{2}$ for $\rho^0$, $c_V=1$ for $\rho^-$ and
$K^{*-}$.

The form factors $f^{P}_{+}(s_{B})$ and $f^{P}_{0}(s_{B})$ given in
Ref. [17] are valid in the full physical range $ 0\leq s_{B} \leq
(1-m_{P})^{2}$. So we will use the  form factors given in Ref. [17]
to estimate the branching ratio of the decay process  $B \to
P\nu\bar{\nu}$, which are same as those used in Ref. [4]. While for
the form factors $V(s_{B})$, $A_1(s_B) $, and $A_2(s_B) $, we will
use those given by Ref. [18], which is same as the form factors (set
C) used in Ref. [4]. It has been shown [4] that the differential
branching ratio for  the decay process  $B \to
K^{\ast}\nu\bar{\nu}$ is similar for sets A and B, there is a
difference of about $ 25\%$ relative to the results obtained from
set C. Certainly, this conclusion also applies to our paper. The
detailed expressions of the form factors $f^{P}_{i}$ and $A_{i}$
 are listed in Appendix.

The di-neutrino invariant mass distributions for the decay processes
$B \to P\nu\bar{\nu}$ and $B \to V\nu\bar{\nu}$ can be written as:
\begin{eqnarray}
\frac{d\mathcal{B}(B \to P\nu_{i'}\bar{\nu}_i)}{ds_B}
&=&\left|C^\nu_L+C^\nu_R\right|^2
\frac{\tau_Bm_B^3}{2^5\pi^3c_P^2}\lambda_P^{3/2}(s_B)\left[f^{P}_+(s_B)\right]^2,\\
\frac{d\mathcal{B}(B \to V\nu_{i'}\bar{\nu}_i)}{ds_B}
&=&\left|C^\nu_L+C^\nu_R\right|^2
\frac{\tau_Bm_B^3}{2^7\pi^3c_V^2}\lambda_V^{1/2}(s_B)
\frac{8s_B\lambda_V(s_B)V^2(s_B)}{(1+\sqrt{r_V})^2}\nonumber\\
&+&\left|C^\nu_L-C^\nu_R\right|^2
\frac{\tau_Bm_B^3}{2^7\pi^3c_V^2}\lambda_V^{1/2}(s_B)\frac{1}{r_V}
\left[(1+\sqrt{r_V})^2(\lambda_V(s_B)+12r_Vs_B)A_1^2(s_B)\frac{}{}\right.\nonumber\\
&+&\left.\frac{\lambda_V^2(s_B)A_2^2(s_B)}{(1+\sqrt{r_V})^2}
-2\lambda_V(s_B)(1-r_V-s_B)A_1(s_B)A_2(s_B)\right].
\end{eqnarray}
Here
\begin{eqnarray}
\lambda_V(s_B)=\lambda(1,r_V,s_B/m_B^2)=1+r_{V}^{2}+\frac{s^{2}}{m^{4}_{B}}-2r_{V}-\frac{2s}{m^{2}_{B}
}-\frac{2r_{V}s_{V}}{m_{B}^{2}},
\end{eqnarray}
where $r_V=m^2_V/m_B^2$, and $\lambda_P$ is similar to $\lambda_V$
by changing $V$ to $P$.

%table
\begin{table}
\begin{center}
\begin{displaymath}
\begin{tabular}{|l|l|}
\hline \spp $G_F = 1.166 \times 10^{-5} \; \gev^{-2}$ &
     $m_{B_d}=5.279\; \gev$ \\
\spp $\alpha = 7.297 \times 10^{-3}$ &
     $m_{B_u}=5.279 \; \gev$  \\
\spp $\tau_{B_u} = (1.638) \times 10^{-12} s$  &
     $V_{tb}= 1.0 $  \\
\spp $\tau_{B_d} = 1.53 \times 10^{-12} s$  &
     $V_{ts}= (40.6 \pm 2.7) \times 10^{-3}$ \\
\spp $M_W= 80.425(38) \; \gev$ & $V_{td}= (9.4\pm3.6)\times 10^{-3}$\\
\spp $sin^2\theta_w=0.23120(15)$ &$m_{t}=175\pm9 \; \gev$
     \\ \hline
\end{tabular}
\end{displaymath}
\caption{Numerical inputs used in our analysis. Unless explicitly
specified, they are \hspace*{1.7cm} taken from the Particle Data
Group [19].\label{tab:inputs}}
\end{center}
\end{table}

To obtain numerical results, we need to specify the relevant $SM$
parameters. Most of these input parameters have been shown in Table
\ref{tab:inputs}. It is obvious that, except these $SM$ input
parameters, the branching ratio $Br(B\rightarrow M\nu\bar{\nu})$ is
dependent on the model dependent parameters $M_{Z'}$ and $K_{1}$.
The lower limits on the mass parameter $M_{Z'}$ predicted by the
topcolor scenario can be obtained via studying its effects on
various observables, which have been precisely measured in the high
energy collider experiments [12]. The most severe constraints come
from the precision electroweak data, which demand that the $Z'$ mass
$M_{Z'}$ must be larger than $1 TeV$ [20]. The vacuum tilting, the
constraints from $Z$-pole physics, and $U(1)$ triviality require
$K_{1}\leq 1$ [21]. Thus, in our numerical calculation, we will take
them as free parameters and assume that they are in the ranges of
$1~\rm TeV\leq M_{Z'}\leq2~\rm TeV$ and $0 < K_{1}\leq 1$.

\begin{table}
\begin{center}
\footnotesize{
\begin{tabular}{c|c|c|cccccc}
\hline\hline Observable & Exp. Data& SM Predictions & $M_{Z'}(\rm GeV)$ =$1100$ & $1400$ &$1700$ & $2000$  \\ \hline
$\mathcal {B}(B_{d}^{0}\rightarrow K^{0}\nu\bar{\nu})$ & $ <160$ & $[3.48  , 6.55] $
&$6.41$&$5.55$  &$5.26$  & $5.15$ \\
$$ & $$ & $$ &  $10.59$ &$ 7.14$&$ 6.00$& $5.53$\\
\hline
$\mathcal {B}(B_{u}^{+}\rightarrow K^{+}\nu\bar{\nu})$  & $<14$ & $[3.75,7.04] $
& $8.19$ & $6.46$ & $5.89$ & $5.66$ \\
$$ & $$ & $$  & $16.56$ &$ 9.65$&$ 7.36$& $6.42$\\
\hline
$\mathcal {B}(B_{d}^{0}\rightarrow \pi^{0}\nu\bar{\nu})$  &$<220$& $[0.05,0.12]  $
&  $0.61$  & $0.29$ & $0.18$ & $0.14$ \\
$$ & $$ & $$ &  $2.18$ &$ 0.89$&$0.46$& $0.28$\\
\hline
$\mathcal {B}(B_{u}^{+}\rightarrow \pi^{+}\nu\bar{\nu})$   & $<100$ & $[0.11,0.25]  $
& $1.22$ & $0.58 $ &$0.36$ & $0.27$\\
$$ & $$ & $$  & $4.37$ &$ 1.78$&$0.91$& $0.56$\\
\hline
$\mathcal {B}(B_{d}^{0}\rightarrow K^{\ast0}\nu\bar{\nu})$  & $<120$& $[6.98,15.19]   $
& $163.09$& $69.02$ & $37.73$ & $25.00$ \\
$$ & $$ & $$ &  $618.45$ &$ 242.56$&$ 117.56$& $66.67$\\
\hline
$\mathcal {B}(B_{u}^{+}\rightarrow K^{\ast+}\nu\bar{\nu})$  & $<80$& $[7.55,16.35]   $
&  $327.43$& $132.18$ & $67.25$ & $40.82$ \\
$$ & $$ & $$ &  $1272.62$ &$ 492.41$&$ 232.94$& $127.31$\\
\hline
$\mathcal {B}(B_{d}^{0}\rightarrow \rho^{\ast0}\nu\bar{\nu})$   & $<440$& $[0.10,0.29]  $
& $72.42$& $27.73 $ & $12.86$ & $6.81$\\
$$ & $$ & $$  & $287.9$ &$ 109.85$&$ 50.63$& $26.52$\\
\hline
$\mathcal {B}(B_{u}^{+}\rightarrow \rho^{\ast+}\nu\bar{\nu})$   & $<150$& $[0.22,0.62]  $
& $144.84$&$55.47$ & $25.74$ & $13.64$\\
$$ & $$ & $$  & $1377.35$ &$525.19$&$ 241.79$& $126.42$\\
\hline\hline
\end{tabular}}
\caption{\small The values (in units of $10^{-6}$ ) of the branching
ratios for the semi-leptonic decays \hspace*{1.7cm}$B\rightarrow
M\nu\bar{\nu}$($M=\pi,K,\rho,K^{\ast}$) for  $ K_{1}=0.4$ (the first
line), $K_{1}=0.8$ (the second line), \hspace*{1.7cm}and  different
values of $M_{Z'}$. }
\end{center}
\end{table}

The values (in units of $10^{-6}$ ) of the branching ratios for the
semi-leptonic decays $B\rightarrow M\nu\bar{\nu}$
($M=\pi,K,\rho,K^{\ast}$), contributed by the nonuniversal gauge
boson $Z'$, are displayed in Table 2 for the coupling parameter $
K_{1}=0.4$ (the first line of every row) and $0.8$ (the second line
of every row). The second and third columns in Table 2 express the
corresponding experimental upper limits and the $SM$ prediction
values, respectively. From this table, one can see that the
nonuniversal gauge boson $Z'$ predicted by the $TC2$ model can
indeed generate significant contributions to these $FCNC$ decay
processes. The values of their branching ratios are sensitive to the
free parameters $M_{Z'}$ and $ K_{1}$. In most of the parameter
space, the contributions of $Z'$ to the $FCNC$ decay processes $B
\to V\nu\bar{\nu}$ are larger than those for the $FCNC$ decay
processes $B \to P\nu\bar{\nu}$, which is easily apprehended from
Eqs. (12) and (13) and the relevant couplings of the nonuniversal
gauge boson $Z'$ with quarks given in Eqs. (1) and (2). In wide
range of the parameter space, the new gauge boson $Z'$ can make the
values of the branching ratio $Br(B \to V\nu\bar{\nu})$ exceed the
corresponding experimental upper limit.

\begin{figure}[htb]
\begin{center}
\epsfig{file=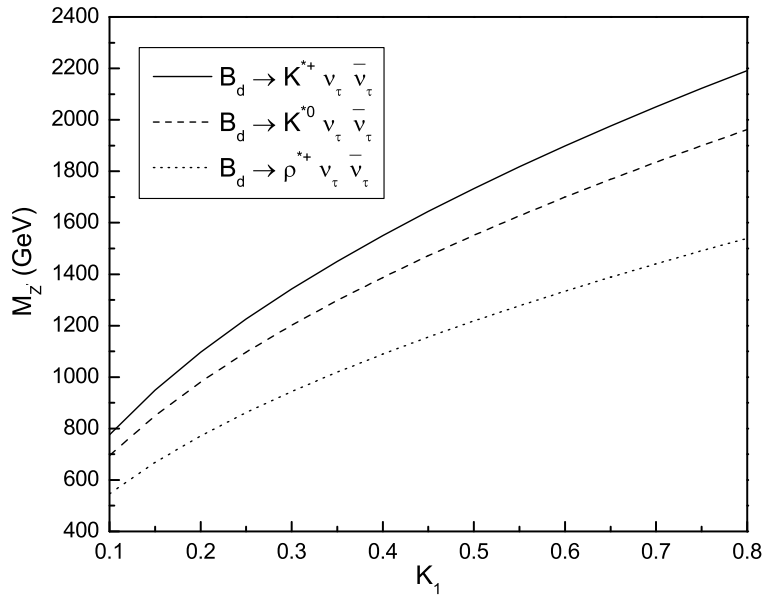,scale=1.1} \caption{The mass parameter
$M_{Z'}$  as a function of the coupling parameter $K_{1}$ for
\hspace*{1.8cm} $Br^{TC2}(B \to V\nu\bar{\nu})= Br^{exp}(B \to
V\nu\bar{\nu})$. }
\end{center}
\end{figure}

To see whether the present experimental upper limit for the
branching ratio $Br(B \to V\nu\bar{\nu})$ can give constraints on
the free parameters of the $TC2$ model, we let that its value equals
to the corresponding experimental upper limit, and plot the mass
parameter $M_{Z'}$ as a function of the coupling parameter $K_{1}$
in Fig.2, in which the solid line, dotted line, and dashed line
denote the $FCNC$ decay processes $B_{u}^{+}\rightarrow
K^{\ast+}\nu_{\tau}\bar{\nu}_{\tau}$,  $B_{d}^{+}\rightarrow
\rho^{\ast+}\nu_{\tau}\bar{\nu}_{\tau}$, and $B_{d}^{0}\rightarrow
K^{\ast0}\nu_{\tau}\bar{\nu}_{\tau}$, respectively. From this
figure, we can see that the present experimental upper limits of
these $FCNC$ decay processes can indeed give severe constraints on
the relevant free parameters. The constraints coming from the $FCNC$
decay process $B_{u}^{+}\rightarrow
K^{\ast+}\nu_{\tau}\bar{\nu}_{\tau}$ is the strongest, which demands
that if we desire $M_{Z'}\leq 2~\rm TeV$, there must be $K_{1}\leq
0.65$.

The presence of the physical scalars, the top-pions
$\pi_{t}^{0,\pm}$ and the top-Higgs boson $h_t^0$, in the low energy
spectrum is an inevitable feature of the topcolor scenario,
regardless of the dynamics responsible for $EWSB$ and other quark
masses [12].  These new particles treat the third generation
fermions differently from those in the first and second generation
fermions and thus can lead to the tree level $FC$ couplings to
ordinary fermions. So they can also generate contributions to some
$FCNC$ processes.

\vspace{1.0cm}
\begin{figure}[htb]
\vspace{2cm}
\begin{center}
\epsfig{file=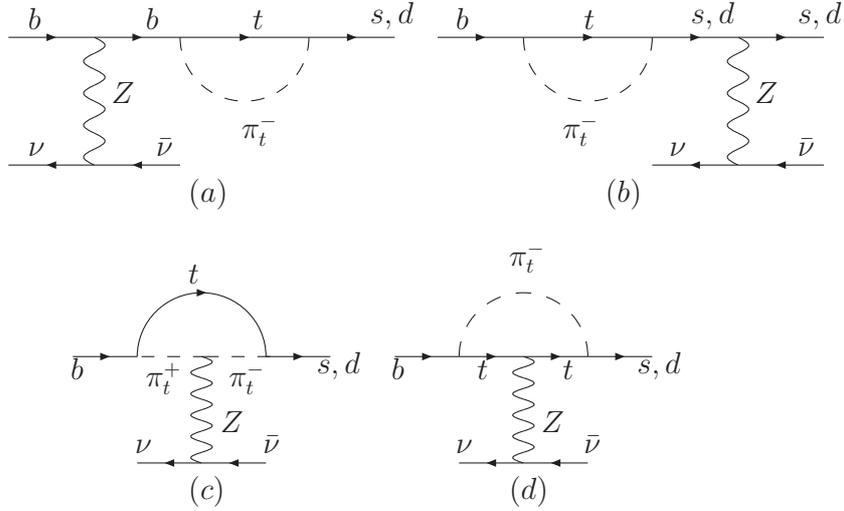,scale=0.8} \caption{The penguin diagrams for
$\pi_{t}^{\pm}$ contributing to the rare decay processes
\hspace*{1.9cm} $b\rightarrow (s,d)\nu\bar{\nu}$ in the $TC2$
model.}
\end{center}
\end{figure}

In the context of the $TC2$ model, the couplings of the charged
top-pions $\pi_{t}^{\pm}$ to ordinary fermions, which are related to
our calculation, can be written as [6, 13, 22]:
\begin{eqnarray}
\frac{m_t^*}{F_{\pi}} \frac{\sqrt{\nu_{w}^{2}-
F_{\pi}^{2}}}{\nu_{w}}\left[\bar{t}_{R}b_{L}
\pi_{t}^{+}+\bar{b}_{L}t_{R}\pi_{t}^{-}+K_{UR}^{tc^{*}}K_{DL}^{ss}\bar{t}_{R}s_{L}\pi_{t}^{+}
+K_{UR}^{tc}K_{DL}^{ss^{*}}\bar{s}_{L}t_{R}\pi_{t}^{-}\right],
\end{eqnarray}
where $m_t^*=m_{t}(1-\varepsilon)$, $\nu_{w}=\nu/\sqrt{2}=174\rm
~GeV$, $F_{\pi}\approx50\rm ~GeV$ is the top-pion decay constant.
$K_{UR}$ and $K_{DL}$ are rotation matrices that diagonalize the
up-quark  and down-quark mass matrices $M_{U}$ and $M_{D}$, i.e.,
$K_{UL}^{+} M_{U}K_{UR}=M_{U}^{dia}$ and
$K_{DL}^{+}M_{D}K_{DR}=M_{D}^{dia}$, for which the $CKM$ matrix is
defined as $V=K_{UL}^{+}K_{DL}$. To yield a realistic form of the
$CKM$ matrix $V$, it has been shown that the values of the coupling
parameters can be taken as [22]:
\begin{eqnarray}
 K_{DL}^{ss}\approx 1, \hspace{10mm}
 K_{UR}^{tc}\leq\sqrt{2\varepsilon-\varepsilon^{2}}.
\end{eqnarray}
In numerical estimation, we will take
$K_{UR}^{tc}=\sqrt{2\varepsilon- \varepsilon^{2}}$ and assume that
the value of the free parameter $ \varepsilon $ is in the range of
$0.03- 0.1$.

The charged top-pions $\pi_{t}^{\pm}$ can contribute to the quark
level transition $b\rightarrow d_{j}\nu\bar{\nu} (j=s,d)$ via the
penguin diagrams, as shown in Fig.3. However, the $FC$ coupling
$\pi_t^{\pm}ts$ or $\pi_t^{\pm}td$ is suppressed by a factor
$K_{UR}^{tc}$ with $\varepsilon$ in the range of $0.03- 0.1$.
Thus, the contributions of the top-pions $\pi^{\pm}_{t}$ to the 
rare decay processes $b\rightarrow (s,d)\nu\bar{\nu}$ are much
smaller than those of the nonuniversal gauge boson $Z'$. Our
numerical results show that it indeed is this case. The value of the
branching ratio $Br(B\rightarrow M \nu\bar{\nu})$ contributed by the
scalars $\pi^{\pm}_{t} $ is smaller than that of $Z'$ at least by
two orders of magnitude, which is consistent with the conclusion
obtained in Ref. [23].

In Ref. [23], we consider the contributions of the $TC2$ model to
the branching ratios and asymmetry observables related to the quark
level transition $b \to sl^+l^-$. We find that the contributions of
the scalar predicted by the $TC2$ model to the decay process $B \to
K \tau^+\tau^-$ are smaller than those of the nonuniversal gauge
boson $Z'$ by two orders of magnitude and therefore can be
neglected. When the $Z'$ mass is in the range of $1000\rm~ GeV-
2000\rm~ GeV$, the value of $Br(B \to K \tau^+\tau^-)$ is in the
range of $7.0\times10^{-6}-1.7\times10^{-6}$, which is larger
than those for the decay process $B \to K e^+e^-$ or $B \to K
\mu^+\mu^-$. This is because of the large coupling of $Z'$ to the
third generation fermions. If we assume that the
 experimental constraint for the branching ratio of the rare decay process
 $B\to K l^{+} l^{-}$ provided by BaBar
 and Belle experiments is $Br(B\rightarrow K l^{+}
l^{-})  = (1.6\pm 0.5)\times10^{-6}$ [24], then we can easily obtain
the constraints on the free parameters $ K_{1}$ and $M_{Z'}$. For
example, for $K_{1} = 0.4$, there must be $ 1290GeV \leq M_{Z'} \leq
1787GeV$. In the case of considering these constraints on the
relevant free parameters, the contributions of the nonuniversal
gauge boson $Z'$ to the rare decays $B\rightarrow M \nu\bar{\nu}$
would be reduced. For instance, for $K_{1} = 0.4$ and $ 1290GeV \leq
M_{Z'} \leq 1787GeV$, there are $2.14\times 10^{-5} \leq
Br(B_{d}^{+}\rightarrow \rho^+\nu_{\tau}\bar{\nu}_{\tau})\leq
7.64\times 10^{-5}$, $5.70\times10^{-5}\leq Br(B_{u}^{+}\rightarrow
K^{\ast+}\nu_{\tau}\bar{\nu}_{\tau}) \leq 1.82\times10^{-4}$, and
 $3.31\times10^{-5}\leq Br(B_{d}^{0}\rightarrow
K^{\ast0}\nu_{\tau}\bar{\nu}_{\tau})\leq 9.11\times10^{-5} $. It is
also possible that the value of the branching ratio $B\rightarrow M
\nu\bar{\nu}$ is larger than the corresponding value predicted by
the $SM$. Thus, the contributions of the nonuniversal gauge boson
$Z'$ to the quark level transition processes $b \to sl^+l^-$ are
correlated with those for the quark level transition processes $b
\to s\nu\bar{\nu}$. However, even if the experimental measurement
value of the branching ratio $Br(B \to X_{s}l^+l^-)$ gives severe
constraints on the relevant free parameters, it is still possible to
largely enhance the branching ratios related to the quark level
transition processes $b \to s\nu\bar{\nu}$ in the $TC2$ model. These
conclusions are consistent with those given in Ref. [4] for a
general $Z'$ model.

\vspace{0.3cm}

\noindent{\bf \large 3. The $LHT$ model and the rare decays
$B\rightarrow M \nu\bar{\nu}$}

Little Higgs theory [8] was proposed as an alternative solution to
the hierarchy problem of the $SM$, which provides a possible kind of
the $EWSB$ mechanism accomplished by a naturally light Higgs boson.
In matter content, the littlest Higgs ($LH$) model [7] is the most
economical little Higgs model discussed in the literature, which has
almost all of the essential feature of the little Higgs models. In
order to make this model consistent with electroweak precision tests
and simultaneously having the new particles of this model in the
reach of the $LHC$, a discrete symmetry, T-parity, has been
introduced, which forms the $LHT$ model [9]. This new physics model
is one of the attractive little Higgs models. In which, all the $SM$
particles are even and among the new particles only a heavy $2/3$
charged T quark belongs to the even sector.

A consistent implementation of T-parity also requires the
introduction of mirror fermions -- one for each quark and lepton
species [9, 25]. The masses of the T-odd fermions can be written in
a unified manner:
\begin{equation}
M_{F_{i}}=\sqrt{2}k_{i}f,
\end{equation}
where $k_{i}$ are the eigenvalues of the mass matrix $k$ and their
values are generally dependent on the fermion species $i$. These new
fermions (T-odd quarks and T-odd leptons) have new flavor violating
interactions with the $SM$ fermions mediated by the new gauge bosons
$(A_{H}$, $W_{H}^{\pm}$, or $Z_{H})$ and at higher order by the
triplet scalar $\Phi$. These interactions are governed by the new
mixing matrices $V_{Hd}$ and $V_{Hl}$ for down-quarks and charged
leptons, respectively. The corresponding matrices in the up quark
($V_{Hu}$) and neutrino ($V_{H\nu}$) sectors are obtained by means
of the relations [9, 26]:
\begin{equation}
V_{Hu}^{+}V_{Hd}=V_{CKM},\hspace*{0.2cm}V_{H\nu}^{+}V_{Hl}=V_{PMNS},
\end{equation}
where the $CKM$ matrix $V_{CKM} $ is defined through flavor mixing
in the down-type quark sector, while the $PMNS$ matrix $V_{PMNS} $
is defined through neutrino mixing.

The details of the $LHT$ model as well as the particle spectrum,
Feynman rules, and its effects on some $FCNC$ processes have been
studied in Ref. [10]. An $O(\upsilon^{2}/f^{2})$ contribution to the
relevant $Z$-penguin diagrams and the corrected Feynman rules of
Ref. [10] are given in Ref. [11].

From the above discussions, we can see that, although the $LHT$
model does not introduce new operators in addition to the $SM$ ones,
it is not minimal flavor violation ($MFV$) because of the mirror
fermions mixing. The mirror fermions introduce a new mechanism for
$FCNC$ processes. Thus, the $LHT$ model might generate significant
contributions to the $FCNC$ processes $B\rightarrow
P(V)\nu\bar{\nu}$  via correcting the coefficient $C^{\nu}_{L,SM}$
given by Eq. (4).
\begin{figure}[htb]
\begin{center}
\epsfig{file=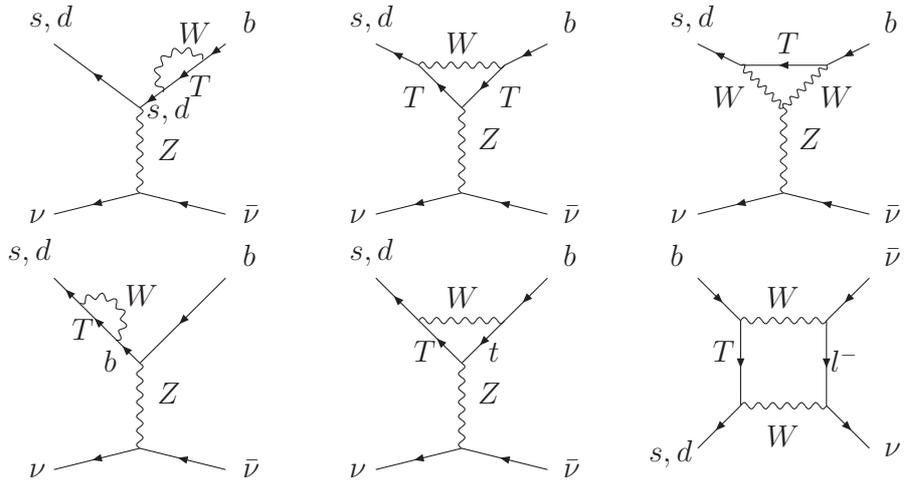,scale=0.8} \caption{The Feynman diagrams of
$T$-even heavy top quark $T$ contributing to the rare
\hspace*{1.7cm} decay processes $b\rightarrow (s,d)\nu\bar{\nu}$ in
the $LHT$ model.}
\end{center}
\end{figure}
\begin{figure}[htb]
\begin{center}
\epsfig{file=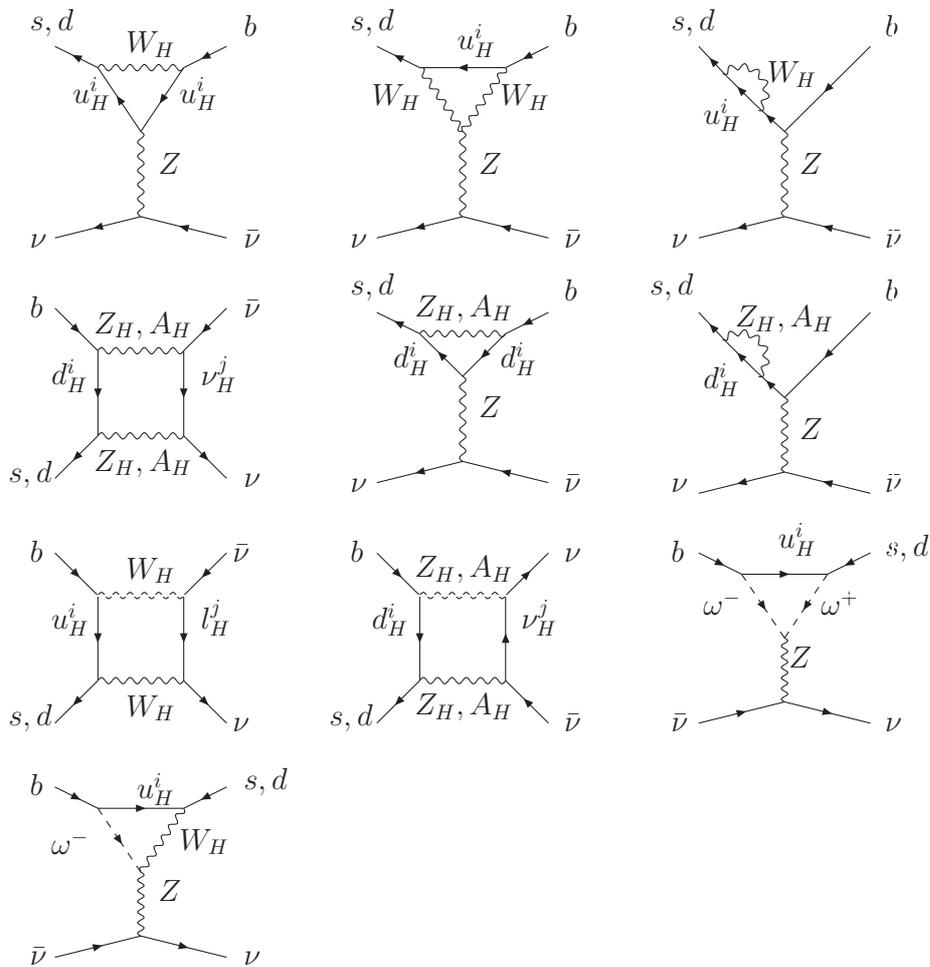,scale=0.8} \caption{The Feynman diagrams of
T-odd fermions contributing to the rare decay \hspace*{1.8cm}
processes $b\rightarrow (s,d)\nu\bar{\nu}$ in the $LHT$ model.}
\end{center}
\end{figure}

The contributions of the $LHT$ model to the quark level transition
processes $b\rightarrow d_{j}\nu\bar{\nu} (j=s,d)$ come from two new
sources: $T$-even heavy top quark $T$ and the $T$-odd fermions,
which can generate contributions to the coefficient
$C^{\nu}_{L,SM}$. The relevant Feynman diagrams are shown in Fig.4
and Fig.5. From the discussions given in section 2, we can see that
the $LHT$ model contributes to the rare decay processes
$B\rightarrow M \nu\bar{\nu}$ ($M=\pi,K,\rho,K^{\ast}$) through the
modification of the function $X_{SM}$ which is related to the
coefficient $C^{\nu}_{L,SM}$.

It is obvious that the branching ratios $Br(B\rightarrow M
\nu\bar{\nu})$ ($M=\pi,K,\rho,K^{\ast}$) contributed by the $LHT$
model are dependent on the free parameters $f$, $x_{L}$, the $T$-odd
fermion masses $M_{Q^{i}_{H}}$, and the  flavor mixing matrix
elements $(V_{H_{d}})_{ij}$. The mixing matrix elements
$(V_{H_{u}})_{ij}$ can be determined via
$V_{H_{u}}=V_{H_{d}}V_{CKM}^{\dag}$. The matrix $V_{H_{d}}$ can be
parameterized in terms of three mixing angles and three phases,
which can be probed by the $FCNC$ processes in $K$ and $B$ meson
systems, as discussed in detail in Refs. [26, 10]. To avoid any
additional parameters introduced and to simplify our calculations,
we take $V_{H_{d}}=V_{CKM}$ and $V_{H_{u}}^{\dag}=I $, and assume
the $T$-odd fermion masses $M_{Q^{i}_{H}}$ in two scenarios:\\
Case I: $M_{Q^{1}_{H}}=M_{Q^{2}_{H}}=M_{Q^{3}_{H}}$.\\
Case II: The $T$-odd fermion masses $M_{Q^{i}_{H}}$ are not
degenerate.

Case I is the $MFV$ limit of the $LHT$ model. In this case, the
contributions of the $T$-odd fermions to the rare decay processes
$B\rightarrow M \nu\bar{\nu}$ ($M=\pi,K,\rho,K^{\ast}$) equal to
zero from the unitarity of the matrix  $V_{H_{d}}$. The
contributions of the $LHT$ model to these $FCNC$ processes are only
coming from the $T$-even heavy top quark $T$, which are dependent on
two parameters $x_{L}$ and $f$. The relative functions are given
by[10, 11]
\begin{eqnarray}
X_{SM}(x_{t})&=&\frac{x_{t}}{8}\left[\frac{x_{t}+2}{x_{t}-1}+
\frac{3x_{t}-6}{(x_{t}-1)^{2}}\log
x_{t}\right],\\
\bar{X}_{even}&=&x^{2}_{L}\frac{\upsilon^{2}}{f^{2}}\left[U_{3}(x_{t},
x_{T})+\frac{x_{L}}{1-x_{L}}\frac{x_{t}}{8}\right],\\
U_{3}(x_{t},x_{T}) &=& \frac{-3+2x_{t}-2x^{2}_{t}}{8(-1+x_{t})}-\frac{x_{t}(-4-x_{t}+2x^{2}_{t})
\log x_{t}}{8(-1+x_{t})^{2}}+\frac{(3+2x_{t})\log x_{T}}{8},
\end{eqnarray}
where the parameters $x_{i}$ are defined as
\begin{eqnarray}
x_{t} &=& \frac{m^{2}_{t}}{M^{2}_{W}},\quad\quad x_{T} =
\frac{m^{2}_{T}}{M^{2}_{W}}.
\end{eqnarray}

For case II, the contributions of the $LHT$ model to the rare decay
processes $B\rightarrow M \nu\bar{\nu}$ come from T-even and T-odd
sectors. The expression forms of the functions $X_{i} $, which are
related to our calculation, can be written as:
\begin{eqnarray}
X_{s,d}&=& X_{SM} + \bar{X}_{even}+\frac{1}{\lambda_{t}^{s,d}}\bar{X}^{odd}_{s,d},
\end{eqnarray}
where the functions $ X_{SM}$ and $\bar{X}_{even} $ have been given
in Eq. (19) and  Eq. (20), respectively, the function
$\bar{X}^{odd}_{s,d}$ is [10, 11]
\begin{eqnarray}
\bar{X}^{odd}_{s,d}&=&\left[\xi^{s,d}_{2}(J^{\nu\bar{\nu}}(z_{2},y)-J^{\nu\bar{\nu}}(z_{1},y))+\xi^{s,d}_{3}
(J^{\nu\bar{\nu}}(z_{3},y)-J^{\nu\bar{\nu}}(z_{1},y))\right]
\end{eqnarray}
with
\begin{eqnarray}
J^{\nu\bar{\nu}}(z_{i},y)&=&\frac{1}{64}\frac{\upsilon^{2}}{f^{2}}\left[z_{i}S_{odd}+F^{\nu\bar{\nu}}(z_{i},y;W_{H})\right.\nonumber\\
&&+\left.4\left(G(z_{i},y;Z_{H})+G_{1}(z'_{i},y';A_{H})+G_{2}(z_{i},y;\eta)\right)\right],\\
S_{odd}&=&\frac{z_{i}^{2}-2z_{i}+4}{(1-z_{i})^{2}}\log
z_{i}+\frac{7-z_{i}}{2(1-z_{i})},\\
\xi^{s,d}_{2}&=&\lambda_{c}^{s,d}, \quad\quad\xi^{s,d}_{3}=\lambda_{t}^{s,d},\\
\lambda_{c}^{s}&=&V^{\ast}_{cb}V_{cs},\quad\lambda_{c}^{d}=V^{\ast}_{cb}V_{cd}, \quad\lambda_{t}^{s}=V^{\ast}_{tb}V_{ts},
\quad\lambda_{t}^{d}=V^{\ast}_{tb}V_{td}.
\end{eqnarray}
Here the functions $F^{\nu\bar{\nu}},G,G_{1}$ and $G_{2}$ given in
Appendix and the various variables defined as follows
\begin{eqnarray}
z_{i}&=&\frac{M^{2}_{Q^{i}_{H}}}{M_{W_H}^{2}}=\frac{M^{2}_{Q^{i}_{H}}}{M_{Z_H}^{2}}, \quad\quad z'_{i}= az_{i},
\quad\quad a=\frac{5}{\tan^{2}\theta_{w}},\\
y&=&\frac{M^{2}_{L_{H}}}{M_{W_H}^{2}}=\frac{M^{2}_{L_{H}}}{M_{Z_H}^{2}},
\quad\quad y'= ay, \quad\quad \eta=\frac{1}{a}\quad .
\end{eqnarray}
The mass of the T-odd heavy gauge boson $W_{H}$ can be written as $
M_{W_{H}} = f g(1-\frac{\upsilon^{2}}{8f^{2}})$ and there is $
M_{W_{H}}\simeq M_{Z_{H}}$.

In the context of the $LHT$ model, the branching ratios of the rare
decays $B\rightarrow X_{s,d}\nu\bar{\nu}$ can be written as:
\begin{eqnarray}
\mathcal{B}(B\rightarrow
X_{s}\nu\bar{\nu}) &=&
\left|\frac{X_{s}}{X_{SM}}\right|^{2}\mathcal{B}(B\rightarrow
X_{s}\nu\bar{\nu})_{SM},\\
\mathcal{B}(B\rightarrow X_{d}\nu\bar{\nu}) &=& \left|\frac{X_{d}}{X_{SM}}\right|^{2}\mathcal{B}(B\rightarrow
X_{d}\nu\bar{\nu})_{SM}.
\end{eqnarray}
\begin{figure}[htb]
\begin{center}
\epsfig{file=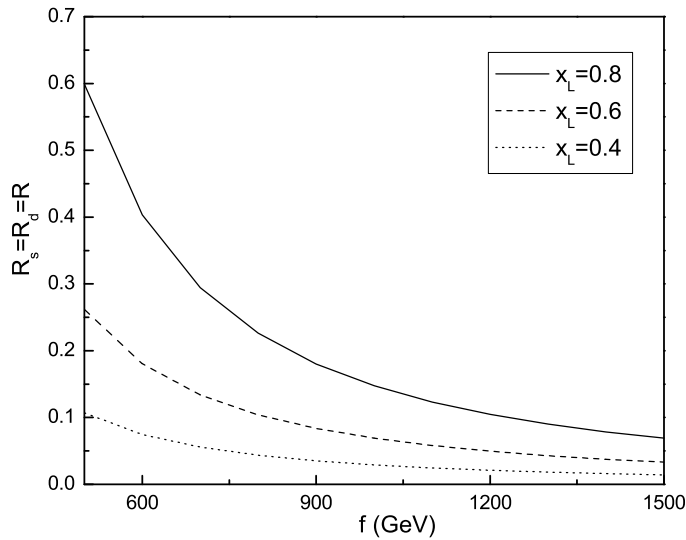,scale=1.0} \caption{The relative correction
parameter $R$ as a function of the scale parameter $f$ for \hspace*{1.7cm}three value of the mixing parameter $ x_{L}$
 in case I.}
\end{center}
\end{figure}
\begin{figure}
\centering \subfigure[$$]{
\includegraphics[scale=0.87]{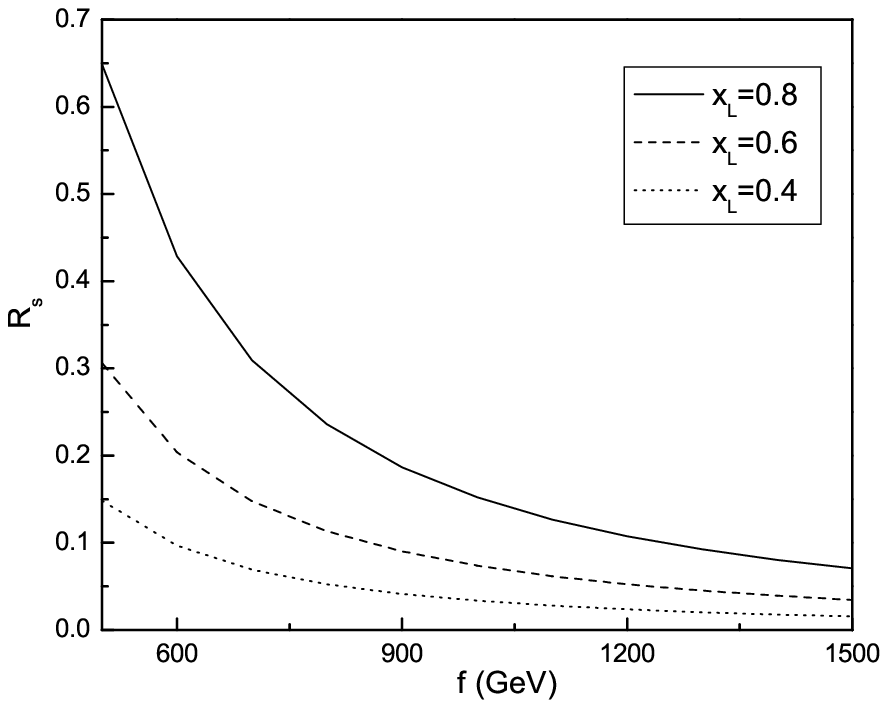}}
%\hspace{1in}
\subfigure[$$]{
\includegraphics[scale=0.87]{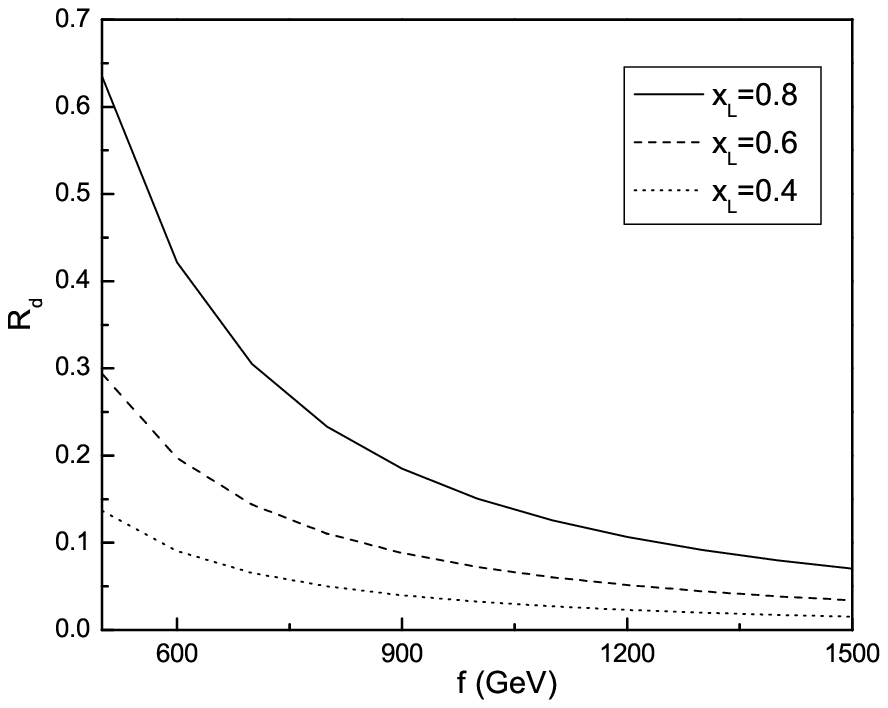}}
\caption{ Same as Fig.6 but for case
II.} \label{fig:9}
\end{figure}
To see the contributions of the $LHT$ model to  the rare decays
$B\rightarrow X_{s,d}\nu\bar{\nu}$, we define the relative
correction parameters $R_{s}$ and $R_{d}$ as
\begin{eqnarray}
R_{s}&=&\frac{\mathcal{B}(B\rightarrow X_{s}\nu\bar{\nu})-\mathcal{B}(B\rightarrow X_{s}\nu\bar{\nu})_{SM}}
{\mathcal{B}(B\rightarrow X_{s}\nu\bar{\nu})_{SM}}=\frac{X_{s}^{2}-X_{SM}^{2}}{X_{SM}^{2}},\\
R_{d}&=&\frac{\mathcal{B}(B\rightarrow X_{d}\nu\bar{\nu})-\mathcal{B}(B\rightarrow X_{d}\nu\bar{\nu})_{SM}}
{\mathcal{B}(B\rightarrow X_{d}\nu\bar{\nu})_{SM}}=\frac{X_{d}^{2}-X_{SM}^{2}}{X_{SM}^{2}}.
\end{eqnarray}

For case I, because the contributions of $T$-odd particles disappear
and the contributions of the $LHT$ model to the rare decay processes
$B\rightarrow M \nu\bar{\nu}$ only come  from the $T$-even heavy top
quark $T$ which are dependent on the free parameters $x_{L}$ and
$f$. If we see these processes at the  quark level, we can obtain
 $X_{s}=X_{d}$ and thus there is $R_{s}=R_{d}$. For case II, both
$T$-even and $T$-odd particles can contribute to these $FCNC$ decay
processes. From Eqs. (24)--(28), we can see that the functions
$X_{s}$ and $X_{d}$ are different from each other due to
$\frac{1}{\lambda_{t}^{s,d}}\bar{X}^{odd}_{s,d}$. Thus, in case II,
 the T-odd fermion masses not being degenerate, there is $R_{s} \neq R_{d}$.

Our numerical results are summarized in Fig.6 and Fig.7 for case I
and case II, respectively. In Fig.6, we have assumed
$R_{s}=R_{d}=R$, in Fig.7 we have taken $M_{Q^{1}_{H}}=700~\rm GeV$,
$M_{Q^{2}_{H}}=1000~\rm GeV$, $M_{Q^{3}_{H}}=1500 ~\rm GeV$ and
$M_{L_{H}}=500~\rm GeV$. One can see from Fig.6 and Fig.7 that the
contributions of the $LHT$ model to the rare $B$ decays
$B\rightarrow M\nu\bar{\nu}(M=\pi, K, \rho, K^{\ast})$ are smaller
than those of the $TC2$ model.  For the scale parameter $f\geq 1TeV
$ and the mixing parameter $ x_{L}\leq 0.8$, the value of the
correction parameter $R_{s}=R_{d}=R$ contributed by the $T$-even
heavy top quark $T$ is smaller than $14.7 \%$, which is consistent
with the numerical result given by Fig.5 of Ref. [10]. In case II,
the $T$-odd particles have contributions  to the rare decay
processes $B\rightarrow M \nu\bar{\nu}$. However, their
contributions are smaller than those of the $T$-even heavy top quark
$T$. For example, for $M_{Q^{1}_{H}}=700~\rm GeV$,
$M_{Q^{2}_{H}}=1000~\rm GeV$, $M_{Q^{3}_{H}}=1500 ~\rm GeV$,
$M_{L_{H}}=500~\rm GeV$, $f \geq 500GeV $, and $x_{L} \leq 0.8$, the
value of the relative correction parameter $R_{s}$ contributed by
the $T$-odd particles is smaller than  $5 \%$. Certainly, in this
paper, we have taken $V_{H_{d}}=V_{CKM}$, which is a very limited
scenario. In more general scenarios, as discussed in Ref. [10], the
contributions of the $T$-odd particles can be enhanced. However, in
most of the parameter space of the $LHT$ model, the value of the
relative correction parameter $R_{s}$ or $R_{d}$ contributed by the
$T$-odd particles is smaller than $10 \%$. It is well known that the
$SM$ prediction values for the branching ratios of the rare $B$
decays $B\rightarrow M\nu\bar{\nu}$ have large uncertainties. Thus,
we have to say that it is very difficult to detect correction
effects of the $LHT$ model on the rare $B$ decays $B\rightarrow
M\nu\bar{\nu}$ in near future high energy collider experiments.

\vspace{0.3cm}

\noindent{\bf \large 4. Conclusions }

The $TC2$ model and the $LHT$ model are two kinds of popular new
physics models beyond the $SM$. The new particles predicted by these
two new physics models can induce $FC$ couplings to ordinary
particles and thus can produce contributions to some $FCNC$
processes. The rare $B$ semileptonic decays with neutrinos in the
final state are significantly suppressed in the $SM$ and their
long-distance contributions are generally subleading. So these
$FCNC$ processes are considered as excellent probes of new physics
beyond the $SM$. In this paper, we consider the contributions of the
$TC2$ model and the $LHT$ model to the rare $B$ decay processes
$B\rightarrow M\nu\bar{\nu}$ with $M=\pi,\rho, K, K^\ast$ and
discuss the possibility of constraining the relevant free parameters
using the corresponding experimental upper limits. The following
conclusions are obtained.

$i)$ The contributions of the $TC2$ model to these rare decay
processes are larger than those from the $LHT$ model. We might use
these processes to distinguish different new physics models in
future high energy collider experiments.

$ii)$ The contributions of the $TC2$ model to these rare decay
processes mainly come from the nonuniversal gauge boson $Z'$. The
contributions of $Z'$ to the quark level transition processes $b \to
sl^+l^-$ are correlated with those for the quark level transition
processes $b \to s\nu\bar{\nu}$. However, even if the experimental
measurement value of the branching ratio $Br(B \to X_{s}l^+l^-)$
gives severe constraints on the relevant free parameters, it is
still possible to largely enhance the branching ratios of the rare
$B$ decay processes $B\rightarrow M\nu\bar{\nu}$ in the $TC2$ model.

$iii)$ The contributions of the nonuniversal gauge boson $Z'$ to the
rare $B$ decays $B\rightarrow V\nu\bar{\nu}$ are larger than those
for the rare $B$ decays $B\rightarrow P\nu\bar{\nu}$. The
experimental upper limits of the branching ratios for some of these
rare decay processes can give constraints on the free parameters of
the $TC2$ model. The most severe constraints on the free parameters
of the $TC2$ model come from the rare $B$ decay
$B^{+}_{u}\rightarrow K^{\ast+}\nu_{\tau}\bar{\nu_{\tau}}$, which
demands that if we desire $M_{Z'}=2~\rm TeV$, there must be
$K_{1}\leq0.5$.

$iv)$ In general, the contributions of the $LHT$ model to the rare
$B$ decays $B\rightarrow M\nu\bar{\nu}$ come from two sources: the
$T$-even and $T$-odd sectors. However, for the case that the $T$-odd
fermions are degenerated in mass, the contributions only come from
the $T$-even heavy top quark $T$. For $f\geq 1TeV $ and $ x_{L}\leq
0.8$, the value of the correction parameter $R$ contributed only by
 $T$ is smaller than $14.7 \%$. In
most of the parameter space of the $LHT$ model, the value of the
relative correction parameter $R_{s}$ or $R_{d}$ contributed by the
$T$-odd particles is smaller than $10 \%$.

\vspace{1.0cm}

\noindent{\bf \large Acknowledgments}

This work is supported in part by the National Natural Science
Foundation of China under Grant No.10975067, the Specialized
Research Fund for the Doctoral Program of Higher Education(SRFDP)
(No.200801650002), the Natural Science Foundation of Liaoning
Science Committee(No.20082148), and the Foundation of Liaoning
Educational Committee(No.2007T086).

\vspace{1.0cm}

\noindent{\bf \large Appendix}

\vspace{0.3cm}

In this appendix we list the functions which are related to our
calculation in the context of the $TC2$ and $LHT$ models. In the
framework of the $TC2$ model:
\begin{eqnarray}
C_{a}(x_{t})&=& \frac{8}{3}(\tan^{2}\theta'-1)\frac{F_{1}(x_{t})}{\upsilon_{d}+ a_{d}},\\
C_{b}(x_{t}) &=& \frac{16F_{2}(x_{t})}{3(\upsilon_{u}-
a_{u})}-\frac{8F_{3}(x_{t})}
{3(\upsilon_{u}+ a_{u})},\\
C_{c}(x_{t}) &=& \frac{16F_{4}(x_{t})}{3(\upsilon_{u}-
a_{u})}+\frac{8F_{5}(x_{t})}{3(\upsilon_{u}+ a_{u})}
\end{eqnarray}
with
\begin{eqnarray}
F_{1}(x_{t}) &=& -(0.5(Q-1)\sin^{2}\theta_{w}+0.25)(x_{t}^{2}\ln(x_{t})/(x_{t}-1)^{2}+x_{t}/(x_{t}-1)\nonumber\\
&-&x_{t}(0.5(-0.5772+\ln(4\pi)-\ln(M^{2}_{W}))+0.75-0.5(x_{t}^{2}\ln(x_{t})/(x_{t}-1)^{2}\nonumber\\
&-&1/(x_{t}-1))))((1+x_{t})/(x_{t}-2)),\\
F_{2}(x_{t}) &=& (0.5Q\sin^{2}\theta_{w}-0.25)(x_{t}^{2}\ln(x_{t})/(x_{t}-1)^{2}-2x_{t}\ln(x_{t})/(x_{t}-1)^{2}\nonumber\\
&+& x_{t}/(x_{t}-1)),\\
F_{3}(x_{t}) &=& -Q \sin^{2}\theta_{w}(x_{t}/(x_{t}-1)-x_{t}\ln(x_{t})/(x_{t}-1)^{2}),\\
F_{4}(x_{t}) &=& 0.25(4\sin^{2}\theta_{w}/3-1)(x^{2}_{t}\ln(x_{t})/(x_{t}-1)^{2}-x_{t}-x_{t}/(x_{t}-1)),\\
F_{5}(x_{t}) &=& -0.25Q\sin^{2}\theta_{w}x_{t}(-0.5772+\ln(4\pi)-\ln(M^{2}_{W})+1-x_{t}\ln(x_{t})/(x_{t}-1))\nonumber\\
&-&\sin^{2}\theta_{w}/6(x^{2}_{t}\ln(x_{t})/(x_{t}-1)^{2}
-x_{t}-x_{t}/(x_{t}-1)).
\end{eqnarray}
Here the variables are defined as: $\upsilon_{u,d}=
I_{3}-2Q_{u,d}\sin^{2}\theta_{w},  a_{u,d}= I_{3}$, where $u$ and
$d$ represent the up- and down-type quarks, respectively. $I_{3}$
is the third component of isospin and $Q_{i}$ is the charge of the
corresponding quark.

The form factors $f^{i}$ for the decay processes $B\rightarrow
(K,\pi)\nu\bar{\nu}$ can be written as [17]:
\begin{eqnarray}
f^{P}_{+}(s_{B})&=& \frac{f(0)}{1-a_{f}(s_{B}/m^{2}_{B})+ b_{f}(s_{B}/m^{2}_{B})^{2}},\\
f^{\pi}(0) &=&0.258\pm0.031, \quad\quad a^{\pi}_{f} = 1.29,  \quad\quad b^{\pi}_{f}= 0.206,\nonumber\\
f^{K}(0)&=& 0.331\pm0.041, \quad\quad a^{K}_{f} = 1.41,  \quad\quad b^{K}_{f} = 0.406\nonumber.
\end{eqnarray}
The form factors for the decay processes $B\rightarrow
(K^{\ast},\rho)\nu\bar{\nu}$ can be written as [18]:
\begin{eqnarray}
A_{i}(s_{B}) &=& A_{i}(0)(1+\beta_{i}s_{B}),\quad\quad \beta_{1}= -0.023 GeV^{-2},\quad\quad \beta_{2}= 0.034 GeV^{-2}.\nonumber\\
V^{B\rightarrow K^{\ast}}(0) &=& 0.411\pm0.033,\quad\quad A^{B\rightarrow K^{\ast}}_{1}(0)= 0.292\pm0.028,\quad\quad
A^{B\rightarrow K^{\ast}}_{2}(0)= 0.259\pm0.025,\nonumber\\
V^{B\rightarrow \rho}(0) &=& 0.323\pm0.030,\quad\quad A^{B\rightarrow \rho}_{1}(0)= 0.242\pm0.023,\quad\quad
A^{B\rightarrow \rho}_{2}(0)= 0.221\pm0.025.\nonumber
\end{eqnarray}

In the $LHT$ model, the relevant functions can be written as [10,
11]:
\begin{eqnarray}
F^{\nu\bar{\nu}}(z_{i},y;W_{H})&=&\frac{3}{2}z_{i}-F_{5}(z_{i},y)-7F_{6}(z_{i},y)-9U(z_{i},y),\\
F_{5}(z_{i},y)&=&\frac{z_{i}^{3}\log z_{i}}{(1-z_{i})(y-z_{i})}+\frac{y^{3}\log y}{(1-y)(z_{i}-y)},\\
F_{6}(z_{i},y)&=&-\left[\frac{z_{i}^{2}\log z_{i}}{(1-z_{i})(y-z_{i})}+\frac{y^{2}\log y}{(1-y)(z_{i}-y)}\right],\\
U(z_{i},y)&=&\frac{z_{i}^{2}\log z_{i}}{(1-z_{i})^{2}(z_{i}-y)}+\frac{y^{2}\log y}{(1-y)^{2}(y-z_{i})}+\frac{1}{(1-z_{i})(1-y)},\\
G(z_{i},y;Z_{H})&=&-\frac{3}{4}U(z_{i},y),\\
G_{1}(z'_{i},y';A_{H})&=&\frac{1}{25a}G(z'_{i},y';A_{H}),\\
G_{2}(z_{i},y;\eta)&=&-\frac{3}{10a}\left[\frac{z_{i}^{2}\log z_{i}}{(1-z_{i})(\eta-z_{i})(z_{i}-y)}\right.\nonumber\\
&&+\left.\frac{y^{2}\log y}{(1-y)(\eta-y)(y-z_{i})}+\frac{\eta^{2}\log\eta}{(1-\eta)(z_{i}-\eta)(\eta-y)}\right].
\end{eqnarray}

\null
\end{document}